\documentclass{ws-procs975x65}

\begin{document}

\title{GENERAL RELATIVISTIC AND NEWTONIAN WHITE DWARFS}

\author{K. BOSHKAYEV$^{1,2,3}$, J.A. RUEDA$^{1,2}$, R. RUFFINI$^{1,2}$ AND I. SIUTSOU$^2$}

\address{$^1$Department of Physics and ICRA, Sapienza University of Rome,\\
Aldo Moro Square 5, I-00185 Rome, Italy\\
$^2$ICRANet, Square of Republic 10, I-65122 Pescara, Italy\\
$^3$Physicotechnical Department, Al-Farabi Kazakh National University,\\
Al-Farabi avenue 71, 050038 Almaty, Kazakhstan;\\
$^*$E-mail: kuantay@icra.it, jorge.rueda@icra.it, ruffini@icra.it, siutsou@icranet.org}

%\author{A. N. AUTHOR}
%
%\address{Group, Laboratory, Street,\\
%City, State ZIP/Zone, Country\\
%E-mail: an\_author@laboratory.com}

\begin{abstract}
The properties of uniformly rotating white dwarfs (RWDs) are analyzed within the framework of Newton's gravity and general relativity. In both cases Hartle's formalism is applied to construct the internal and external solutions to the field equations. The white dwarf (WD) matter is described by the Chandrasekhar equation of state. The region of stability of RWDs is constructed taking into account the mass-shedding limit, inverse $\beta$-decay instability, and the boundary established by the turning points of constant angular momentum $J$ sequences which separates stable from secularly unstable configurations. We found the minimum rotation period $\sim0.28$ s in both cases and maximum rotating masses $\sim1.534 M_{\odot}$ and $\sim1.516 M_{\odot}$ for the Newtonian and general relativistic WDs, respectively. By using the turning point method we show that general relativistic WDs can indeed be axisymmetrically unstable whereas the Newtonian WDs are stable. 

\end{abstract}


\bodymatter

\section{Introduction}\label{aba:sec1}

Recently, equilibrium configurations of non-rotating (static) $^{4}$He, $^{12}$C, $^{16}$O and $^{56}$Fe white dwarfs (WDs) within general relativity (GR) have been constructed in Ref.~\refcite{RotD2011}. The white dwarf matter has been there described by the relativistic generalization of the Feynman-Metropolis-Teller (RFMT) equation of state (EOS) obtained by Rotondo et al.\cite{RotC2011}. A new mass-radius relation that generalizes both the works of Chandrasekhar\cite{Chandra1931} and Hamada \& Salpeter\cite{hamada61} has been there obtained, leading to a smaller maximum mass and a larger minimum radius with respect to the previous calculations. In addition, it has been shown how both GR and inverse $\beta$-decay are relevant for the determination of the maximum stable mass of non-rotating WDs. 

It is therefore of interest to generalize the above results to the case of rotation. As a first attempt, we constructed in Ref.~\refcite{boshCuba} general relativistic uniformly rotating WDs in the simplified case when microscopic Coulomb screening is neglected in the EOS, following the Chandrasekhar\cite{Chandra1931} approximation by describing the matter as a locally uniform fluid of electrons and nuclei. The average molecular weight in the Chandrasekhar EOS is $\mu=A/Z$, where $A$ is the mass number and $Z$ is the number of protons in a nucleus.

As a second attempt, in Ref.~\refcite{boshLH} we calculated the maximum mass of rotating $^{4}$He, $^{12}$C, $^{16}$O and $^{56}$Fe WDs using the Salpeter\cite{salpeter61} and the RFMT EOS. As a result we obtained there different maximum mass for different chemical composition of WD matter. 

As a third attempt, in Ref.~\refcite{boshGX3} we investigated the stability of general relativistic uniformly rotating $^{4}$He, $^{12}$C, $^{16}$O and $^{56}$Fe white dwarfs against secular and dynamical instabilities. We determine the maximum mass and minimum rotation period of stable white dwarfs depending on chemical composition of the white dwarf matter taking into account the Coulomb interactions as well as the nuclear interactions and the electroweak equilibrium at high densities, within the Salpeter\cite{salpeter61} and relativistic Feynman-Metropolis-Teller EOS\cite{RotD2011}.

We summarized all the above results in Ref.~\refcite{boshApJ}, where we analyzed in detail the stability of RWDs from both the microscopic and macroscopic point of view. Besides the inverse $\beta$-decay instability, we also study the limits to the matter density imposed by zero-temperature pycnonuclear fusion reactions using up-to-date theoretical models\cite{Gasques2005, Yakovlev2006}.

In this work we show the importance of general relativity to study the stability of RWDs.  Moreover one can restrict to the Newtonian case when estimating the minimum period of uniformly RWDs. In case of the maximum rotating mass and in the analyses of stability, general relativity indeed becomes relevant.  We make use of Hartle's formalism\cite{H1967,HT1968} for the construction of rotating stars both in Newton's gravity and GR. As a WD matter we consider the Chandrasekhar EOS with average molecular weight, $\mu=2$.  

The main parameters characterizing RWDs such as the total mass $M$, angular momentum $J$, quadrupole moment $Q$, equatorial radius $R_{eq}$ and eccentricity $e$ are obtained for a given EOS, from the matching procedure between the internal and external solutions of the field equations and the equations of hydrostatic equilibrium\cite{H1967,HT1968}. Note, that the total mass is defined by $M=M^{J\neq0}=M^{J=0}+\delta M$, where $M^{J=0}=M^{(0)}$ is the mass of a static white dwarf with the same central density as $M^{J\neq0}$, and $\delta M$ is the contribution to the mass due to rotation.

The main objective of this work is to compare and contrast WDs both rotating and non-rotating in Newton's gravity and general relativity.

\section{Stability criteria}
There are two main stability criteria for non-rotating WDs: the well-known general relativity (GR) instability and inverse $\beta$-decay instability. The inverse $\beta$-decay instability is related to the fact that with the increasing mass of WDs the central density will increase, hence the electrons will become ultra relativistic and their energy will increase until it reaches a certain value of the Fermi energy (depending on chemical composition of a WD matter) which allows the electrons to be captured by the protons inside the  nuclei, forming neutrons. This process is called inverse $\beta$-decay or neutronization and is written in a compact form $p+e^{-}\rightarrow n+\nu_e$, where $p$ is the proton, $e^{-}$ is the electron, $n$ is the neutron and $\nu_e$ is the electron neutrino. Such a process conserves the number of baryons creating neutron rich nuclei through $e^{-}$ capture. Due to the fact that most of the pressure of the WD is supported by the electrons, the inverse $\beta$-decay process leads to the pressure decrease of the WD by increasing the density which causes the instability of the star.

For rotating WDs there exist additional stability criteria. One of them is the mass shedding limit. When a star spins so rapidly the centrifugal forces will prevail over gravitational force  and the star will start losing its matter on the equatorial plane \cite{stergioulas}. The procedure of estimating the mass shedding angular velocity in general relativity was developed in Friedman et. al \cite{Friedman1986}.

In addition to the mass shedding limit there are two more stability criteria called secular (axisymmetric) and dynamical instabilities. Usually in the literature the conditions $T/W=0.14,~e=0.81267$ and $T/W=0.25,~e=0.95288$ are adopted as the upper limit for the onset of the secular and dynamical instabilities\cite{Chandra1969}, where $T/W$ is the rotational/gravitational binding energy and $e$ is the eccentricity of rotating objects. These values are absolute upper limits for the ideal case of Maclaurin spheroids \cite{Chandra1969} and in practice, they cannot be applied to objects like uniformly rotating WDs since they have smaller $T/W$ and $e$ with respect to the Maclaurin spheroids\cite{boshGX3}. However Friedman et. al.\cite{Friedman1988} developed the turning point method of axisymmetric (secular) instability for uniformly rotating stars. The main idea of the turning point method is that along a sequence of constant angular momentum, the condition $(\partial M(\rho,J)/\partial \rho)_{J}=0$ separates secularly stable and unstable stars. For $J=0$ we have the condition for the maximum mass of non-rotating WDs $\partial M(\rho)/\partial \rho=0$. Moreover it was shown in Friedman et. al.\cite{Friedman1988} that a sequence of uniformly rotating stars becomes dynamically unstable after it has become secularly unstable. That is, for fixed angular momentum the point of secular instability occurs at lower central density with respect to dynamical instability.

%%%%%%%%%%%%%%%%%%%%%%%%%%%%%%%%%%%%%%%%%%%%%%%%%%%%%%%%%%%%%%%%%%%%%%%%%%%%%%%%%%%%%%%%%%%%%%%%%%%%%%%%%%%%%%%%%%%%
\begin{figure*}
\centering
\begin{tabular}{lr}
\hspace{-0.5cm}\includegraphics[width=7 cm, height=4.4 cm,clip]{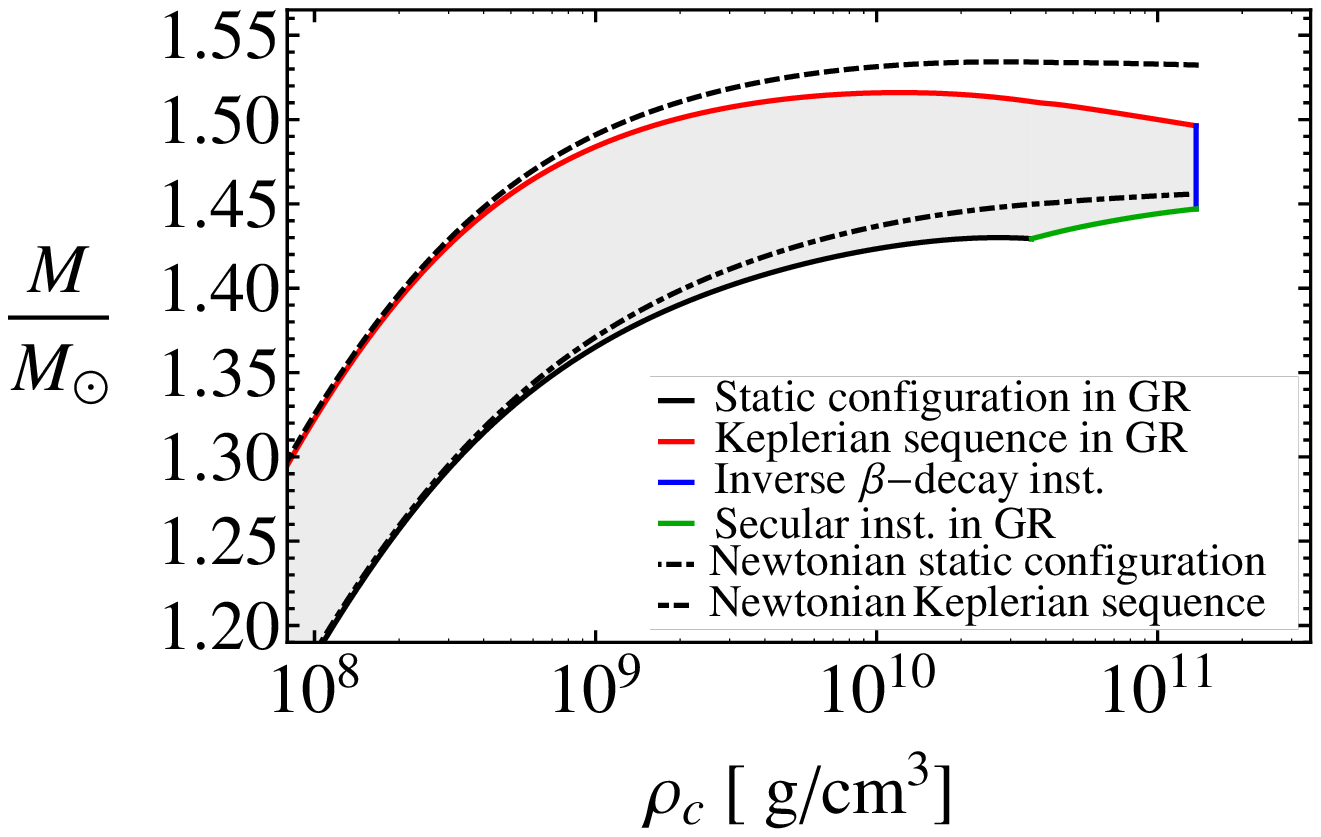} & \hspace{-0.7cm}\includegraphics[width=7 cm, height=4.4 cm,clip]{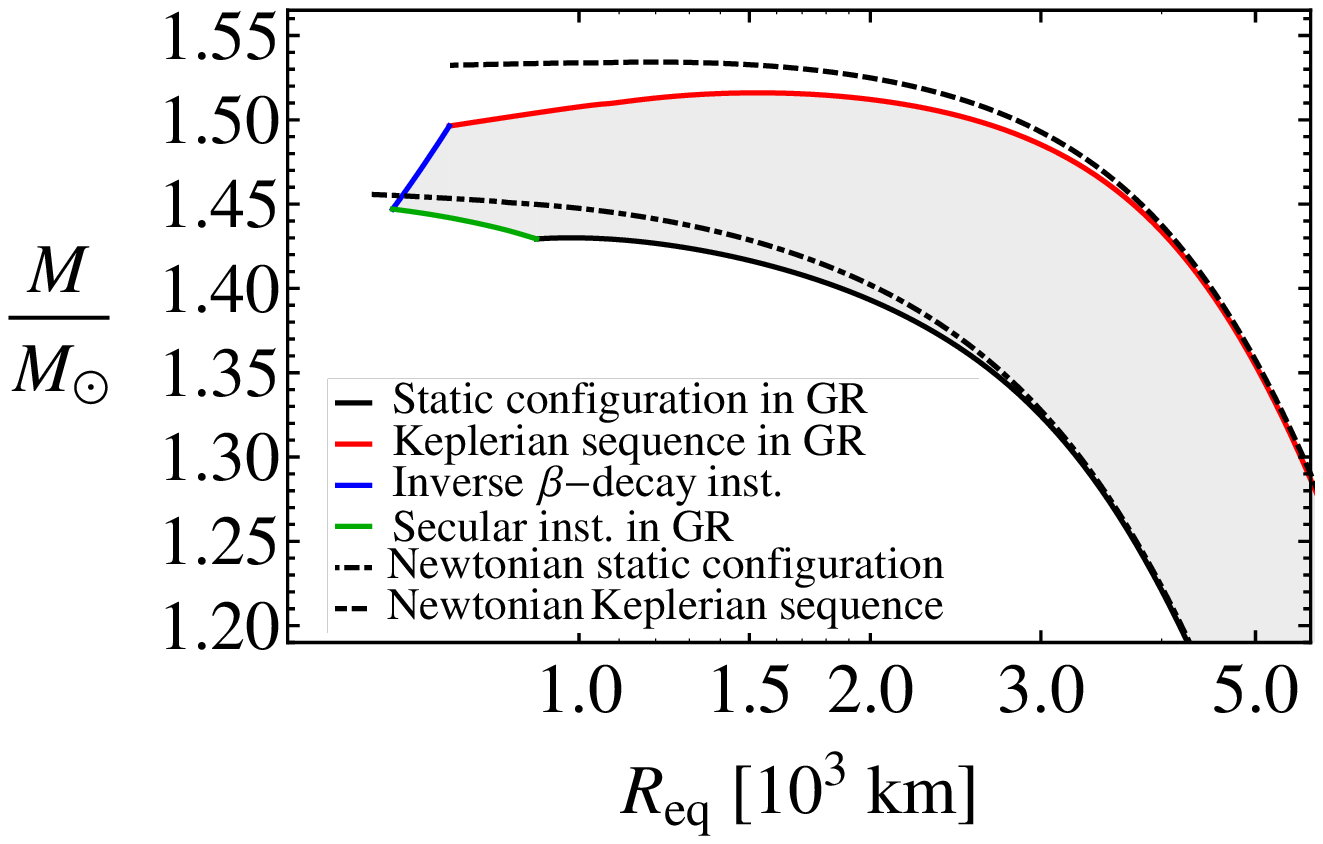}
\end{tabular}
\caption{Mass in solar mass units versus central density (left panel) and mass in solar mass units versus equatorial radius (right panel)  of Newtonian and general relativistic WDs for the Chandrasekhar EOS with $\mu$ = 2. Both the non-rotating case and the Keplerian sequence are shown. We have stopped the density, just for the sake of comparison, at the critical density for the onset of inverse $\beta$-decay of $^4$He = 1.39$\times$10$^{11}$ g/cm$^{3}$.}\label{fig:grnpmrhoreq}
\end{figure*}
%%%%%%%%%%%%%%%%%%%%%%%%%%%%%%%%%%%%%%%%%%%%%%%%%%%%%%%%%%%%%%%%%%%%%%%%%%%%%%%%%%%%%%%%%%%%%%%%%%%%%%%%%%%%%%%%%%%%
\section{Results}
We found that along the Keplerian (the mass shedding) sequence there are no maxima of $J=$constant sequences. This means that along the Keplerian sequence all uniformly rotating WDs are secularly, hence dynamically stable. For this reason we increase the central density all the way up to inverse $\beta$-decay density in order to define the minimum rotation period of WDs, since this is the only limitation which guarantees the stability before the trigger of gravitational collapse. Thus, the minimum period of WDs is determined along the Keplerian sequence at the critical central density for the inverse $\beta$-decay instability.  

%%%%%%%%%%%%%%%%%%%%%%%%%%%%%%%%%%%%%%%%%%%%%%%%%%%%%%%%%%%%%%%%%%%%%%%%%%%%%%%%%%%%%%%%%%%%%%%%%%%%%%%%%%%%%%%%%%%%
\begin{figure*}
\centering
\begin{tabular}{lr}
\hspace{-0.5cm}\includegraphics[width=7 cm, height=4.4 cm, clip]{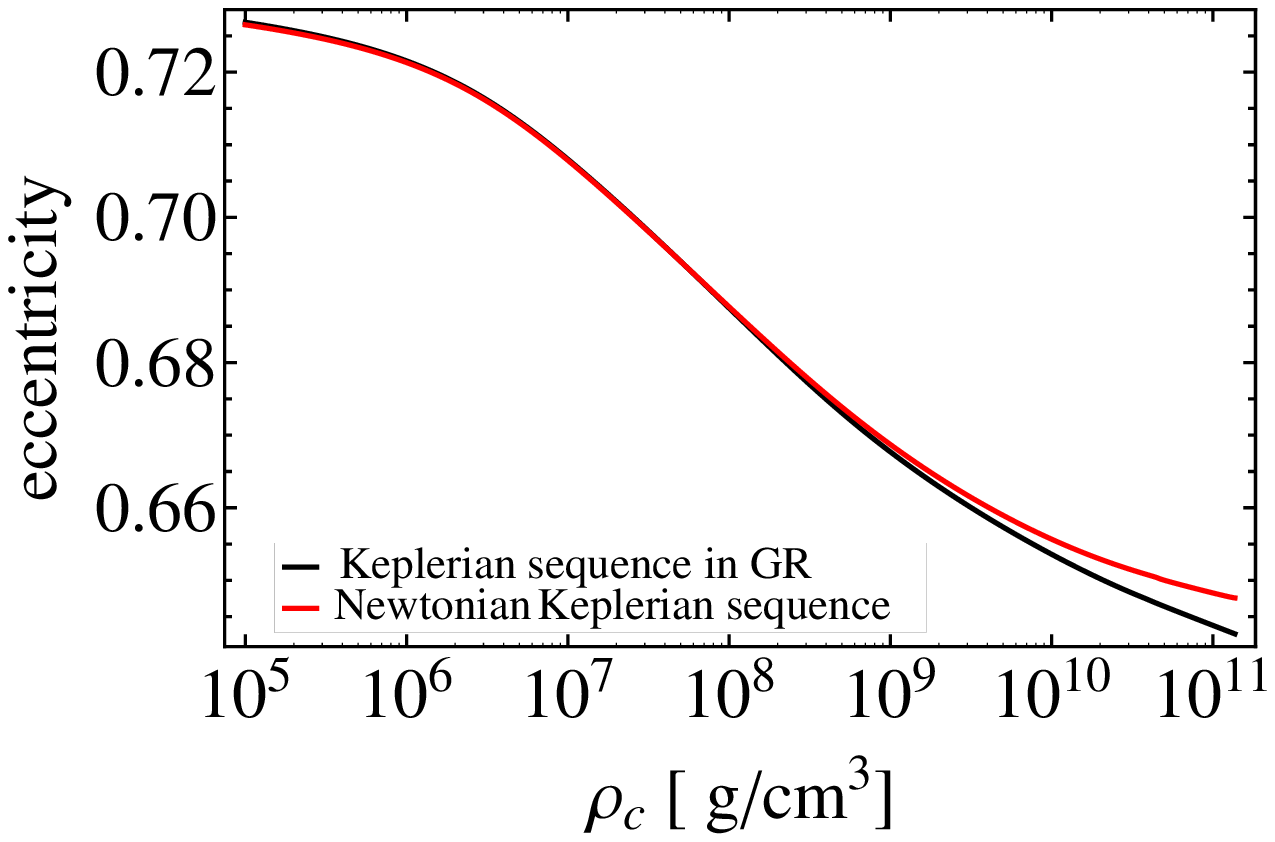} & \hspace{-0.7cm}\includegraphics[width=7 cm, height=4.4 cm, clip]{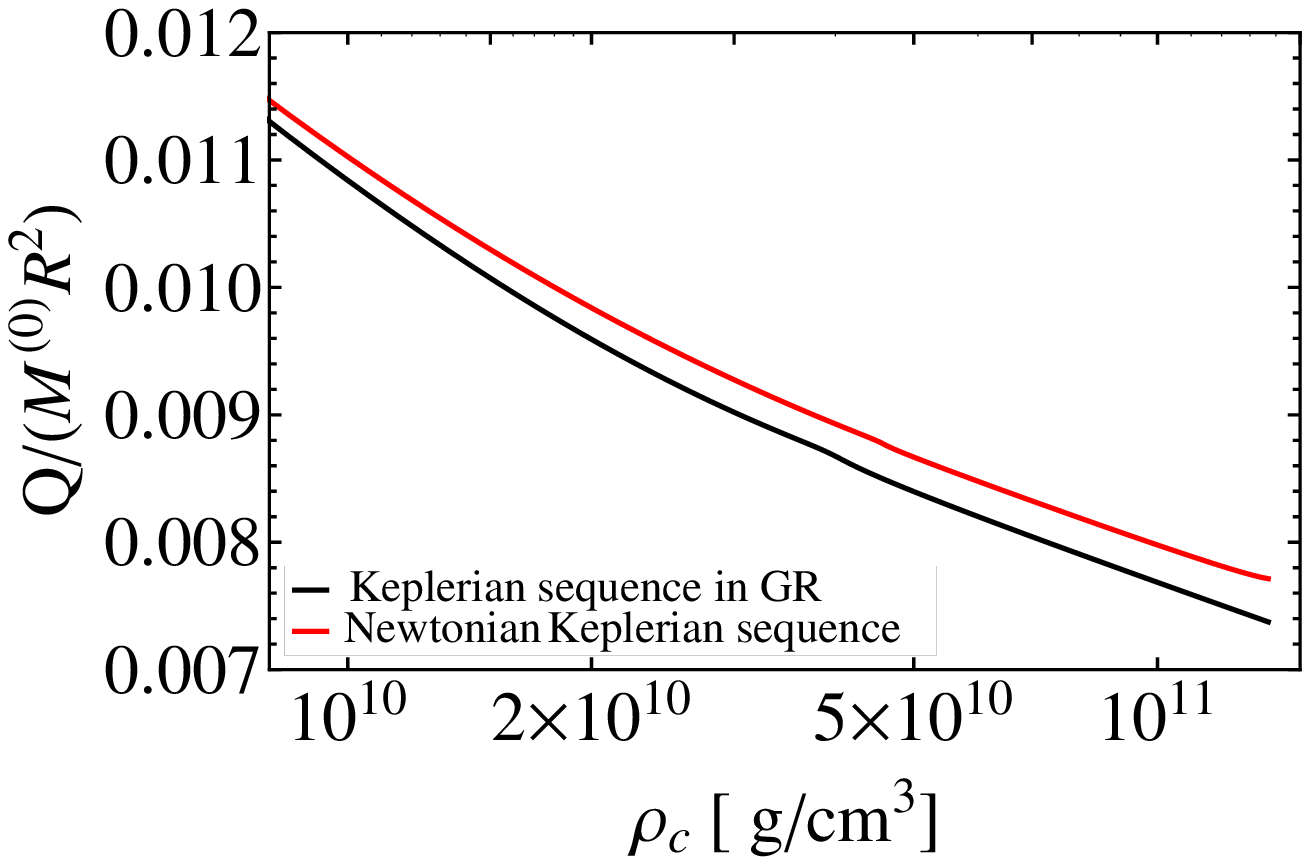}
\end{tabular}
\caption{Left panel: eccentricity $e$ versus  central density $\rho_c$. Right panel:  normalized quadrupole moment $Q/(M^{(0)}R^2)$ versus  versus  central density $\rho_c$, where $M^{(0)}$ and $R$ are the static mass and static radius, respectively. The Keplerian sequences are considered only. The solid red line is in Newton's gravity and the solid black line is in general relativity.}\label{fig:eccq}
\end{figure*}
%%%%%%%%%%%%%%%%%%%%%%%%%%%%%%%%%%%%%%%%%%%%%%%%%%%%%%%%%%%%%%%%%%%%%%%%%%%%%%%%%%%%%%%%%%%%%%%%%%%%%%%%%%%%%%%%%%%%

%%%%%%%%%%%%%%%%%%%%%%%%%%%%%%%%%%%%%%%%%%%%%%%%%%%%%%%%%%%%%%%%%%%%%%%%%%%%%%%%%%%%%%%%%%%%%%%%%%%%%%%%%%%%%%%%%%%%
\begin{figure*}
\centering
\begin{tabular}{lr}
\hspace{-0.5cm}\includegraphics[width=7 cm, height=4.4 cm, clip]{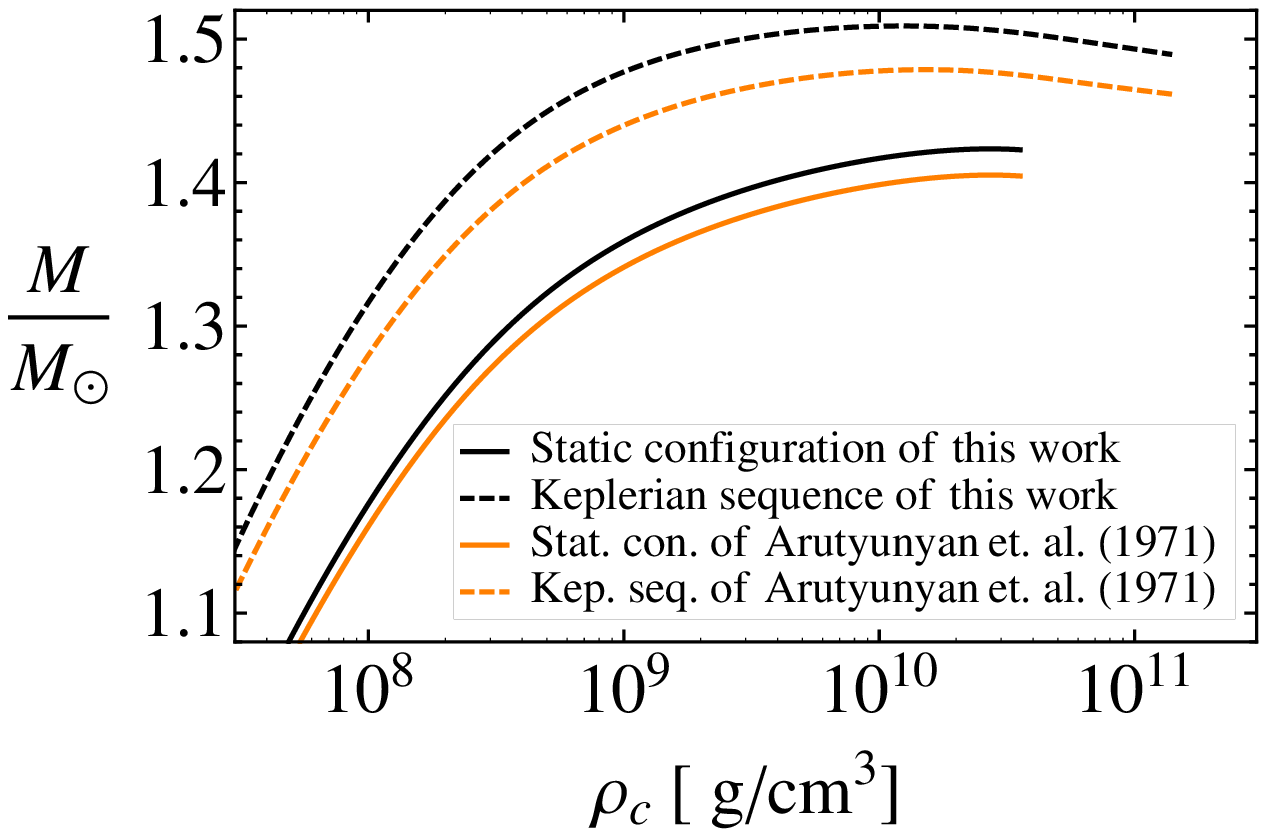} & \hspace{-0.7cm}\includegraphics[width=7 cm, height=4.4 cm, clip]{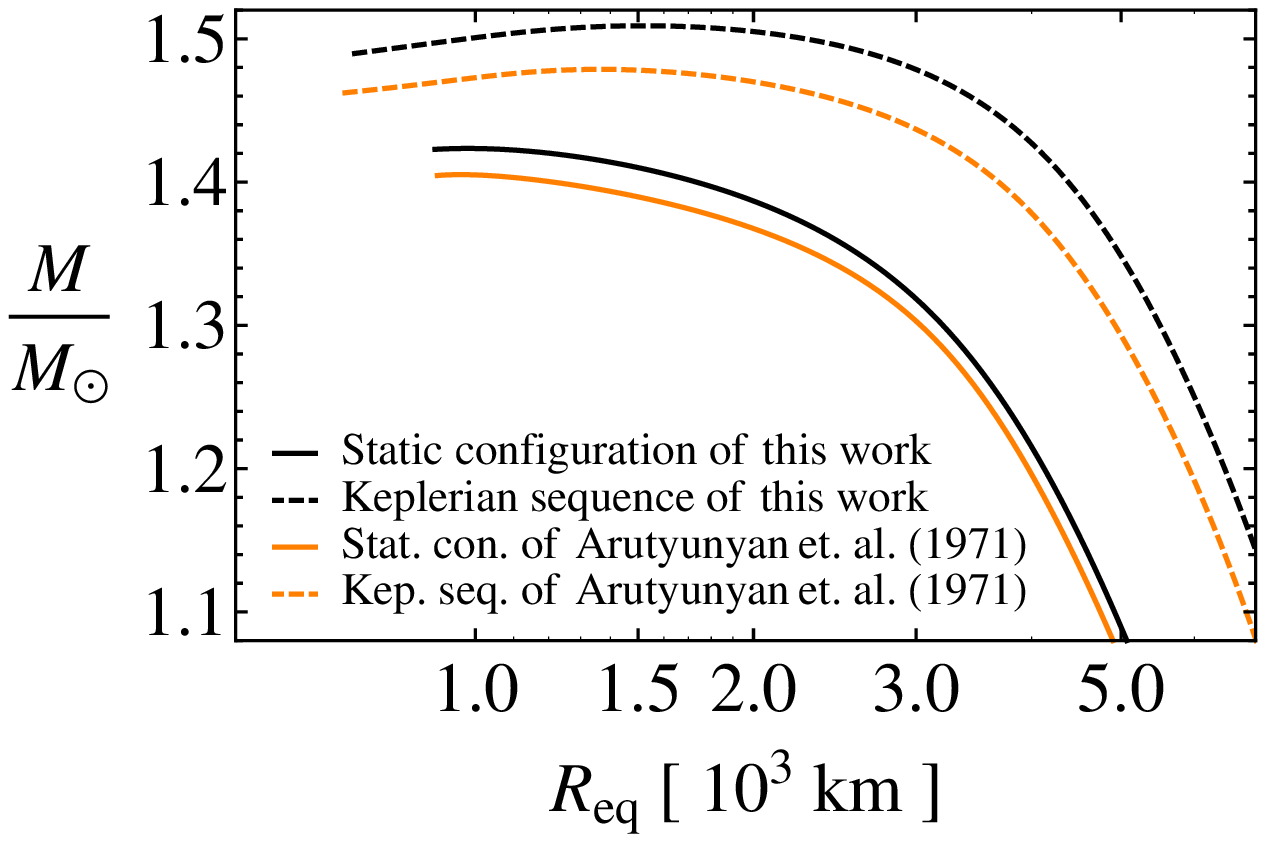}
\end{tabular}
\caption{Mass versus central density relation (left panel) and mass versus equatorial radius for general relativistic WDs using the Chandrasekhar EOS with $\mu$= 2 for the static and the Keplerian sequence of this work and that of Arutyunyan et al. (1971).}\label{fig:arut}
\end{figure*}
%%%%%%%%%%%%%%%%%%%%%%%%%%%%%%%%%%%%%%%%%%%%%%%%%%%%%%%%%%%%%%%%%%%%%%%%%%%%%%%%%%%%%%%%%%%%%%%%%%%%%%%%%%%%%%%%%%%%

In Fig.~\ref{fig:grnpmrhoreq} the mass versus the central density (left panel) and the mass versus the equatorial radius (right panel) are shown for both Newtonian and GR WDs using Hartle's formalism. As we can see in GR the maximum mass is reached at finite central density and its value is smaller than the Newtonian counterpart. Due to this fact in GR RWDs are subjected to the axisymmetric secular instabilities. Unlike GR WDs, Newtonian WDs reach their maximum mass at $\rho_c\rightarrow\infty$, hence there is no axisymmetric instability in Newtonian WDs. The shaded region for GR WDs is the zone where uniformly rotating stable RWDs exist. One can also draw the stability region for the Newtonian WDs. Then this region will be limited from above by the mass shedding (the Keplerian) sequence and from below by the static sequence. Although in principle there is no limit from the right for the Chandrasekhar EOS because it depends only on the average composition $\mu$ and therefore no self-consistent inverse $\beta$-decay instability line can be constructed for different chemical compositions, we assume here to guide the eye the critical inverse $\beta$-decay density of $^4$He, $1.39\times 10^{11}$ g cm$^{-3}$. Heavier compositions have lower critical densities.

Fig.~\ref{fig:eccq} shows the eccentricity versus the central density (left panel) and the normalized quadrupole moment versus the central density (right panel). With the increasing central density the eccentricity and quadrupole moment start to decrease, which shows that the system becomes less oblate. As one can see owing to the general relativistic effects WDs in GR have smaller eccentricity and quadrupole moment than the Newtonian counterparts.

Up to our knowledge, the only previous work on RWDs within GR is the one of Arutyunyan et. al.\cite{1971Ap......7..274A}. A method to compute RWDs configurations accurate up to second order in the angular velocity of a star $\Omega$ was developed by two of the authors (see Ref.~\refcite{sedrakyan1968} for details), independently of the work of Hartle(1967.)\cite{H1967} In Arutyunyan et. al.,\cite{1971Ap......7..274A} RWDs were computed for the Chandrasekhar EOS with $\mu=2$.

In Fig.~\ref{fig:arut}  we show the mass-central density relation (left panel) and the mass-radius relation (right panel) obtained with their method with the ones constructed in this work for the same EOS. We note here that the results are different even at the level of static configurations, and since the methods are based on construction of rotating configurations from seed static ones, those differences extrapolate to the corresponding rotating objects. This fact is to be added to the possible additional difference arising from the different way of approaching the order $\Omega^2$ in the approximation scheme. The differences between the two equilibrium configurations are evident.

\begin{table}

\centering
\tbl{Maximum rotating mass of WDs in literature.}
{\footnotesize
{\begin{tabular}{c c c c}
%
%\toprule
%%
\hline
Treatment/EOS & $M^{J\neq0}_{max}/M_{\odot}$ & References \\
\hline
Newtonian/Chandrasekhar $\mu=2$          &  1.474      &      Anand (1965)\cite{1965PNAS...54...23A} \\
%\hline
%%
%\hline
Newtonian/Polytrope $n=3$         &  1.487      &     Roxburgh (1965)\cite{1965ZA.....62..134R}     \\ 
%\hline
%%
%\hline
Post-Newtonian/Chandrasekhar $\mu=2$     & 1.482      & Roxburgh and Durney (1966)\cite{1966ZA.....64..504R} \\ 
%\hline
%%
%\hline
GR/Chandrasekhar $\mu=2$                & 1.478      &  Arutyunyan et al. (1971)\cite{1971Ap......7..274A} \\
\hline
%%
%\botrule
\end{tabular}\label{tab:mrototh}}
 }
\end{table}

Turning now to the problem of the maximum mass of a RWD, in Table \ref{tab:mrototh} we present the previous results obtained in Newtonian, Post-Newtonian approach and GR by several authors. Depending on their method, approach, treatment, theory and numerical codes the authors showed different results. These maximum masses of RWDs are to be compared with the ones found in this work using the Chandrasekhar EOS $\mu=2$. In Newton's gravity the maximum rotating mass is $1.534M_{\odot}$ and in GR is $1.516M_{\odot}$.

The minimum rotation period of WDs is obtained for a configuration rotating at the Keplerian angular velocity, at the critical inverse $\beta$-decay density; i.e., this is the configuration lying at the crossing point between the mass-shedding and inverse $\beta$-decay boundaries, see Fig.~\ref{fig:grnpmrhoreq}. We found $0.28$ s for both Newtonian and general relativistic white dwarfs. It should be stressed that up to our knowledge the minimum rotation period has never been estimated for uniformly rotating white dwarfs. However Arutyunyan et al. (1971) showed their results for a range of central densities covering the range of interest of our analysis. Thus, we have interpolated their numerical values of the rotation period of WDs in the Keplerian sequence and calculated the precise value at the inverse $\beta$-decay threshold for $\mu$ = 2. We thus obtained minimum period $\sim$0.31 s in agreement with our results.

\section{Conclusions}
We have constructed solutions of the Newtonian equilibrium equations for RWDs accurate up to order $\Omega^2$, following the procedure of Hartle(1967)\cite{H1967}. In Figs.~\ref{fig:grnpmrhoreq},~\ref{fig:eccq} we have compared these Newtonian configurations with general relativistic RWDs for the Chandrasekhar EOS with $\mu=2$. We can see clearly the differences between the two mass-density relations toward the high density region, as expected. The most remarkable difference is the existence of axisymmetric instability boundary in the general relativistic case, absent in its Newtonian counterpart.

We have shown, in particular, that the maximum mass of rotating WDs is stable against secular and dynamical instabilities and determined the minimum rotation period of rotating WDs both for the Newtonian and general relativistic RWDs. We have compared and contrasted the main differences between the Newtonian and general relativistic RWDs and we have shown that Newton's gravity is not enough to properly investigate the stability and to calculate the maximum masses (rotating and non-rotating) of WDs.

The results presented here open the way to a more general description of a rotating WD and are relevant both for the theory of delayed type Ia supernova explosions\cite{Ilkov} as well as for the white dwarf model of Soft Gamma-Ray Repeaters and Anomalous X-Ray Pulsars\cite{M2012}.

\end{document}